\documentclass[aps,preprint]{revtex4-1}
\usepackage{graphicx}
\usepackage{amssymb}
\usepackage{amsmath}
\usepackage{color}
\usepackage{enumerate}
\usepackage{bm}
\begin{document}


\title{Tailoring Heat Dissipation in Ordered Arrays of Dipolar Coupled Magnetic Nanoparticles}	

\author{Manish Anand}
\email{itsanand121@gmail.com}
\affiliation{Department of Physics, Bihar National College, Patna University, Patna-800004, India.}

\date{\today}

\begin{abstract}
The main aim of the present work is to analyse the effect of dipolar interaction strength $\lambda$, particle size $D$ and temperature $T$ on the hysteresis mechanism in ordered arrays of magnetic nanoparticles (MNPs) using computer simulations. The anisotropy axes of the MNPs are assumed to have random orientation to mimic the real system. 
In the absence of thermal fluctuations and dipolar interaction, the hysteresis follows the Stoner and Wohlfarth model irrespective of $D$, as expected. 
The hysteresis loop area is minimal for particle sizes $D \approx8-16$ nm at $T=300$ K and $\lambda=0.0$, indicating the dominance of superparamagnetic character. 
Switching magnetic interaction on is able to move the MNPs from superparamagnetic to a ferromagnetic state even at room temperature; therefore, magnetic interaction of enough strength enhances the hysteresis loop area. Interestingly,
the hysteresis loop area is significant and is the same as that of Stoner and Wohlfarth particle even  $T=300$ K and negligible dipolar interaction for ferromagnetic MNPs ($D>16$ nm). 
The coercive field $\mu^{}_oH^{}_c$ and blocking temperature $T^{}_B$ also get enhanced with an increase in $\lambda$ and $D$. 
The rigorous analysis of the coercive field $\mu^{}_oH^{}_c$ vs temperature data also reveals significant deviation from  $T^{3/4}$ dependence of $\mu^{}_oH^{}_c$ because of dipolar interaction.  
The amount of heat dissipated $E^{}_H$ and $\mu_oH^{}_c$ decrease rapidly with $T$ for $D\approx 8-16$ nm and $\lambda\leq0.6$. On the other hand, $E^{}_H$ and $\mu^{}_oH^{}_c$ depend weakly on $T$ with $D>16$ nm,  even in the weak dipolar limit. 
The present work should provide a better understanding of magnetic hyperthermia to researchers working on this subject. For physicists, it would be interesting to test experimentally the results described in this article.
\end{abstract}
\maketitle
\section{Introduction}


Recent years have evidenced a surge in magnetic nanoparticles (MNPs) based research due to their numerous technological applications~\cite{bader2006,gloag2019,koplovitz2019,akbarzadeh2012,guimaraes2009,gu2016,patra2018,anand2018,Aldewachi2018,anand12021}. 
 MNPs have also received significant attention due to their application in magnetic hyperthermia~\cite{berry2019,moise2018}. In  such a case, the malignant cells are exposed to nanoparticles, and an alternating magnetic field is applied. Due to hysteresis, they release heat which is used to burn the cancerous cells~\cite{moise2018}. The heating efficiency of these nanosystems depends critically on various factors, such as particle size and its distribution, anisotropy constant, magnetic interaction, frequency and strength of the external magnetic field, etc.~\cite{martinez2013,perigo2015,abu2020}. 
Therefore, there is a growing interest to understand how heat release from nanoparticles can be enhanced and manipulated in a more controlled manner.






The various aspects of magnetic hyperthermia are well understood in the case of non-interacting MNPs~\cite{rosensweig2002,raikher2014}. However, MNPs interact due to dipolar interaction. Therefore, their heating capability is drastically modified even for low concentrations of nanoparticles~\cite{ruta2015}. The dipolar interaction is long-ranged
and anisotropic. Depending on the relative orientation and position of the particles, it can favour ferromagnetic or antiferromagnetic interaction~\cite{de1997}. As a consequence, it has a far-reaching effect on various magnetic properties of crucial importance such as modified energy barriers, frustrations of spins, changed magnetic dynamics, unique morphology, etc.~\cite{iglesias2004,plumer2010,anand2016,usov2018,anand2021}. For instance, in the case of randomly distributed particles and anisotropy axes, the magnetic properties of the system can have similarities to those of spin glasses~\cite{djurberg1997}. In other cases, if the MNPs form a linear chain, the dipolar interaction favours head to the tail arrangement of magnetic moments~\cite{barrett2011}.
In these contexts, there are several works which have incorporated the effect of dipolar interaction~\cite{vargas2005,souza2019,allia1999,tan2014,fu2018}. For example, using experiments and analytical methods, Allia {\it et al.} studied the dipolar interaction effect on hysteresis in granular magnetic systems~\cite{allia1999}. 
Dipolar interaction was found to determine the magnetic hysteresis in the high-temperature limit. 
Tan {\it et al.} probed the heating mechanism in the isotropic assembly of MNPs using kinetic Monte Carlo (kMC) simulation~\cite{tan2014}. They obtained spatial distribution of heat dissipation due to dipolar interaction. 
Fu {\it et al.} also studied the dipolar interaction effect on the heating efficiency of MNPs~\cite{fu2018} using computer simulations and experiments. They found  the detrimental effect of magnetic interaction on the heat dissipation in the case of isotropic assembly of MNPs.


In general, superparamagnetic nanoparticles are used for magnetic hyperthermia due to their easy controllability, superior magnetic and physicochemical properties~\cite{ha2018,bao2011}. 
Although, the major disadvantage associated with such particles is their agglomeration, which reduces their dispersibility and hence their intrinsic stability for a more extended period~\cite{chantrell1982,lima2012}. 
The amount of heat dissipation is significantly reduced due to the dipolar interaction in the isotropic assembly of these ultrafine nanoparticles~\cite{fu2018,branquinho2013,usov2017,guibert2015}. 
Usov {\it et al.} analysed the heating performance in a three-dimensional assembly of nanoparticles~\cite{usov2017}. The dipolar interaction lowered the heat dissipation significantly. Using experiments, Guibert {\it et al.} have also shown that the dipolar interaction is detrimental to the heating efficiency of MNPs~\cite{guibert2015}. While in the case of highly anisotropic assembly such as linear chain and columnar geometry; the dipolar interaction improves the heating efficiency of the nanoparticles~\cite{serantes2014,valdes2020,mehdaoui2013,anand2020}. For example, Mehdaoui {\it et al.} found that dipolar interaction improved the heating performance in a columnar arrangement of MNPs~\cite{mehdaoui2013}.
In a recent study, using kMC simulations, we have shown that dipolar interaction improves the heating capability of nanoparticles in a linear chain with perfectly aligned anisotropy~\cite{anand2020}.





The arrangement of MNPs in a particular shape depends on various factors such as dipolar interaction, temperature, etc.~\cite{anderson2016,chuan2012}.
Therefore, we should find a way to maximize the heat dissipation and reduces the adverse effect of dipolar interaction even in the case of isotropic assembly. In this context, there are few recent works which indicate that ferromagnetic nanoparticles can be an excellent alternative to the superparamagnetic counterpart as former has higher heating capability~\cite{kita2008,abu2019,mehdaoui2011,carrey2011,le2016}. For instance, using experiments and numerical simulations, Mehdaoui {\it et al.} 
investigated the heating performance of an assembly of MNPs. Ferromagnetic particles were found to be the best candidate for hyperthermia as they displayed the highest specific losses reported in the literature~\cite{mehdaoui2011}. 
Carrey {\it et al.} also obtained a more substantial heating efficiency of ferromagnetic nanoparticles~\cite{carrey2011}. 
Using experiments, Chang {\it et  al.} studied the heating efficiency in human-like collagen protein-coated MNPs~\cite{le2016}. They also found that ferromagnetic nanoparticles are a better-suited candidate for magnetic hyperthermia. 
Thus motivated, we extensively investigate the hysteresis mechanism in ordered arrays of nanoparticles of various sizes ranging from superparamagnetic to ferromagnetic using kinetic Monte Carlo (kMC) simulation. We also probe the dependence of hysteresis on dipolar interaction strength and temperature. The one-dimensional array is chosen as it is the building block of a variety of possible geometries. We have assumed random orientations of anisotropy axes. kMC simulation is able to simulate the thermodynamic and non-equilibrium processes more efficiently~\cite{ruta2015,tan2014,anand2019}.


The organization of the paper is as follows. The model used in the simulation and various energy terms are discussed in Sec.~II.
We present and discuss the numerical results in Sec.~III.
Finally, we provide the summary and conclusion of the work Sec.~IV







\section{Model}
\begin{figure}[!htb]
\centering\includegraphics[scale=0.50]{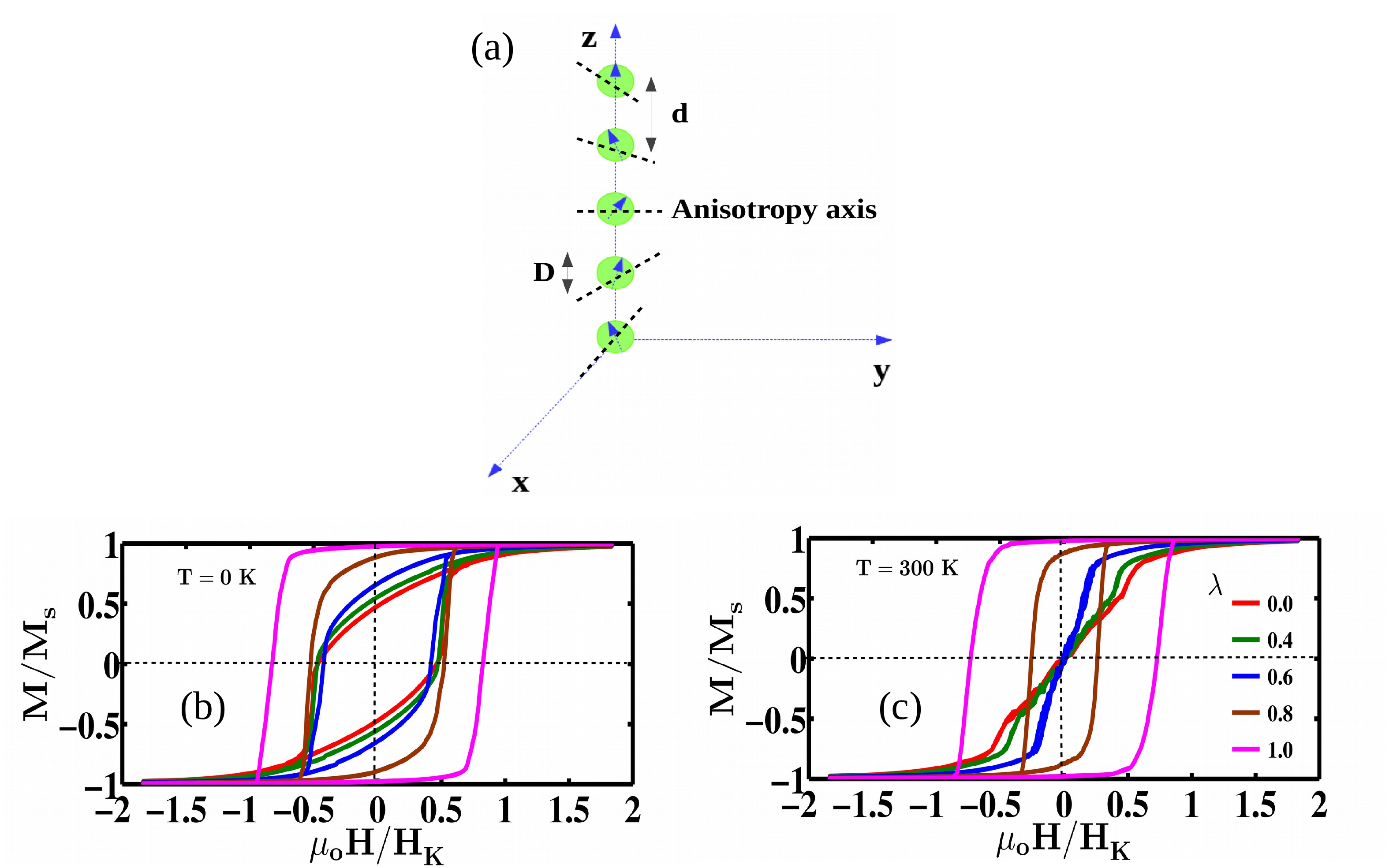}
\caption{(a) Schematic of the linear array of magnetic nanoparticles. The moments are shown with blue coloured vectors while black dashed lines denote the anisotropy axes. The particle has diameter $D$ and $d$ is the distance between two nearest neighbouring particles. Magnetic hysteresis curve as a function of dipolar coupling strength $\lambda$ for superparamagnetic nanoparticle ($D=8$ nm) with two values of temperature (b) $T=0$ K and (c) $T=300$ K. When MNPs are non-interacting ($\lambda=0.0$), the hysteresis curve follows the Stoner and Wohlfarth model in the absence of thermal fluctuations as expected.
The hysteresis loop area is extremely small with $T=300$ K and $\lambda\leq0.6$, indicating superparamagnetic behaviour. There is an increase in the area under the hysteresis curve with dipolar interaction strength $\lambda$ due to enhancement in ferromagnetic coupling.} 
\label{figure1}
\end{figure}

We have considered a linear arrangement of $n$ single-domain particles without any positional disorder as depicted in Fig.~\ref{figure1}(a). The particles are monodisperse and spherical shaped with volume V=$\pi D^3/6$, where $D$ is the diameter of the particle.
Each nanoparticle has a magnetic moment $\mu^{}=\mu\hat{\mu}^{}$ with $\mu=M^{}_sV$, where $\hat{\mu}$ is the unit vector corresponding to $\mu$, and $M^{}_s$ is the saturation magnetization. 
 To vary the dipolar interaction strength, a control parameter $\lambda=D/d$ is defined, $d$ is the separation between two nearest neighbours in the array, as shown in the schematic Fig.~\ref{figure1}(a). The MNPs are non-interacting with $\lambda=0.0$. On the other hand, the strength of the dipolar interaction is the maximum for $\lambda=1.0$.




The energy associated with the single superparamagnetic nanoparticle because of uniaxial anisotropy is given as~\cite{carrey2011,tan2014,anand2019}
\begin{equation}
E=K_{\mathrm {eff}}V\sin^2\theta,
\label{anisotropy}
\end{equation}
here $K_{\mathrm {eff}}$ is the anisotropy constant, and $\theta$ is the angle between the magnetic moment and anisotropy axis.
It is clear that $E$ has minima $E^{}_1$ and $E^{}_2$ positioned at $\theta=0$ and $\pi$, respectively [Eq.~(\ref{anisotropy})]. There is also an energy maximum of strength $E^{}_3=K_{\mathrm {eff}}V$ at $\theta=\pi/2$, also known as the energy barrier. If thermal energy $k^{}_BT$ is strong enough as compared to $K_{\mathrm {eff}}V$, the direction of magnetic moment fluctuates between $E^{}_1$ and $E^{}_2$ with a characteristic time scale, also known as N\'eel relaxation time $\tau^{}_N$. 
It is related to the energy barrier and thermal energy as~\cite{anand2019,manish2020}
\begin{equation}
\tau^{}_N=\tau^{}_o\exp(K_\mathrm {eff} V/k^{}_BT),
\end{equation}  
where $\tau_o=(2\nu^{}_o)^{-1}$ with $\nu^{}_o=10^{10}$ $s^{-1}$, $k^{}_B$ is the Boltzmann constant and $T$ is the temperature. When $\tau^{}_N$ becomes comparable to the experimental measuring time $\tau^{}_m$, i.e. $\tau^{}_N\approx \tau^{}_m$, the nanoparticle is said to be in the blocked state. In this case, the magnetic behaviour of the MNPs is characterized by blocking temperature $T^{}_B$ below which the magnetic moments appear frozen. It is approximated as~\cite{bedanta2008} 
\begin{equation}
T^{}_B\approx\frac{K_{\mathrm {eff}} V}{k^{}_B \ln(\tau_m/\tau_o)},
\label{blocking}
\end{equation}
where $\tau^{}_m\approx(2\pi\nu)^{-1}$. Eq.~(\ref{blocking}) is suitable for  non-interacting particles with the same size and anisotropy.
In an assembly, MNPs interact because of dipolar interaction. As a consequence, the single-particle energy barrier is significantly modified. The dipolar field associated with this interaction is given by the following expression~\cite{tan2014,anand2019}    
\begin{equation}
\mu^{}_{o}\vec{H}^{}_{\mathrm {dip}}=\frac{\mu\mu_{o}}{4\pi}\sum_{j,j\neq i}\frac{3(\hat{\mu}^{}_j \cdot \hat{e}_{ij})\hat{e}_{ij}-\hat{\mu^{}_j} }{r^3_{ij}}.
\label{dipolar}
\end{equation}
Here $\mu^{}_o$ is the vacuum permeability, $\hat{\mu}^{}_j$ is the unit vector for the $j\mathrm{th}$ $\mu$ in the underlying system, $r^{}_{ij}$ is the distance between the two nanoparticles at $i\mathrm{th}$ and $j\mathrm{th}$ positions in the array, and $\hat{e}_{ij}$ is the unit vector corresponding to $r^{}_{ij}$. To obtain the hysteresis response of the underlying system, we apply an oscillating magnetic field ${\mu^{}_{o}H}$  along the array-axis of MNPs as shown below
\begin{equation}
\mu^{}_{o}H=\mu^{}_oH^{}_{\mathrm {o}}\cos\omega t,
\label{magnetic}
\end{equation} 
where $\mu^{}_{o}H_{\mathrm {o}}$ and $\omega=2\pi\nu$ are the magnitude and angular frequency of the external magnetic field, respectively, and $t$ is the time. In the presence of long-ranged dipolar field and externally applied magnetic field, the total energy of the $i\mathrm{th}$ magnetic nanoparticle 
is given by the following expression~\cite{tan2014,anand2019}
\begin{equation}
E(\theta^{}_i,\phi^{}_i)=K_{\mathrm {eff}}V\sin^2 \theta^{}_i-\mu\mu^{}_{o}H^{}_{\mathrm {total}}\cos(\theta^{}_i-\phi^{}_i).
\end{equation}
Here $\phi^{}_i$ is the angle between the anisotropy field and the total
magnetic field $ H^{}_{\mathrm {total}}$ (dipolar and external field).

We have implemented the kMC simulation technique to simulate the hysteresis curves as a function of dipolar interaction strength and the temperature for the particle sizes
ranging from superparamagnetic to ferromagnetic. We have used the same kMC algorithm, which is explained in the references~\cite{tan2014,anand2019}. We have calculated heat released because of hysteresis using the following formula~\cite{carrey2011}



\begin{equation}
E^{}_H=\int \displaylimits_{-\mu^{}_{o}H_{\mathrm{o}}}^{\mu^{}_{o}H^{}_{\mathrm{o}}}M(H)dH,
\label{heat_dissipation}
\end{equation}
where $M(H)$ is the magnetic field dependent magnetization of the uderlying system.

\section{Simulation results}
We have used the following values of parameters: $D=8-64$ nm, $K^{}_{\mathrm {eff}}=13\times10^3$ Jm$^{-1}$, $M^{}_s=4.77\times10^5$ Am$^{-1}$, $\nu=10^3$ Hz and $T=0-300$ K. We have also taken three values of magnetic field strength $\mu^{}_oH^{}_{\mathrm {o}}= 0.05$, 0.075  and 0.1 T, which is about 0.92, 1.38 and 1.84 times the anisotropy field $H^{}_K=2K^{}_{\mathrm {eff}}/M^{}_s$, respectively. These values of the magnetic field and frequency meet the criteria $\mu^{}_oH^{}_{\mathrm {o}} \nu<5\times10^{9}$ Am$^{-1}$s$^{-1}$, a safer limit suited for magnetic hyperthermia applications~\cite{he2018}.
The total number of nanoparticles in the array is taken as $n = 100$, and dipolar interaction strength $\lambda$ has been changed between 0.0 and 1.0. 
$K^{}_{\mathrm {eff}}$ and $M^{}_s$ value considered in the present work is for the bulk magnetite (Fe$_3$O$_4$), which is one of the most promising candidates for magnetic hyperthermia applications due to its bio-compatibility and driving accumulations features~\cite{hachani2017}.


We first study the hysteresis behaviour and its dependence on dipolar interaction strength and temperature for a particle of size well within the superparamagnetic regime.
In Fig.~(\ref{figure1}), we study the hysteresis behaviour as a function of dipolar interaction strength with $T=0$ and 300 K for $D=8$ nm. We have considered five representative values of $\lambda$= 0.0, 0.4, 0.6, 0.8, and 1.0.
The externally applied field axis (x-axis) has been rescaled by $H^{}_K$ and $M$ (y-axis) by $M^{}_s$ in the hysteresis figures shown in this article.
In the absence of thermal fluctuations and magnetic interaction ($\lambda=0.0$), the hysteresis curve resembles the Stoner and Wohlfarth model~\cite{stoner1948}. In this case, coercive field $\mu^{}_oH^{}_c\approx0.479H^{}_K$ and remanent magnetization $M^{}_r$ is close to 0.5, as expected for Stoner and Wohlfarth particle with randomly oriented anisotropy axis~\cite{stoner1948}. The hysteresis loop area is extremely small for weakly or non-interacting MNPs with $T=300$ K indicating superparamagnetic behaviour. It can be explained from the fact that particle size being in the superparamagnetic regime, the blocking temperature $T^{}_B$ is minimal $\approx 16.86$ K [using Eq.~(\ref{blocking})] as compared to simulation temperature $T$ ($=300$ K). Therefore, thermal energy overcomes the energy barrier, instigating the magnetic moments to change their orientations randomly, which results in extremely weak hysteresis.
By solving Fokker–Planck equation, Usov and Grebenshchikov also obtained minimal hysteresis loop area for Cobalt particle in the superparamagnetic regime at $T=250$ K~\cite{usov2009}.
Due to enhancement in the ferromagnetic coupling between the magnetic moments with an increase in dipolar interaction strength, the hysteresis loop area increases with $\lambda$, irrespective of temperature. Mohapatra {\it et al.} also obtained higher hysteresis loss for the linear array as compared to isotropic assembly using experiments~\cite{mohapatra2020}.

\begin{figure}[!htb]
\centering\includegraphics[scale=0.50]{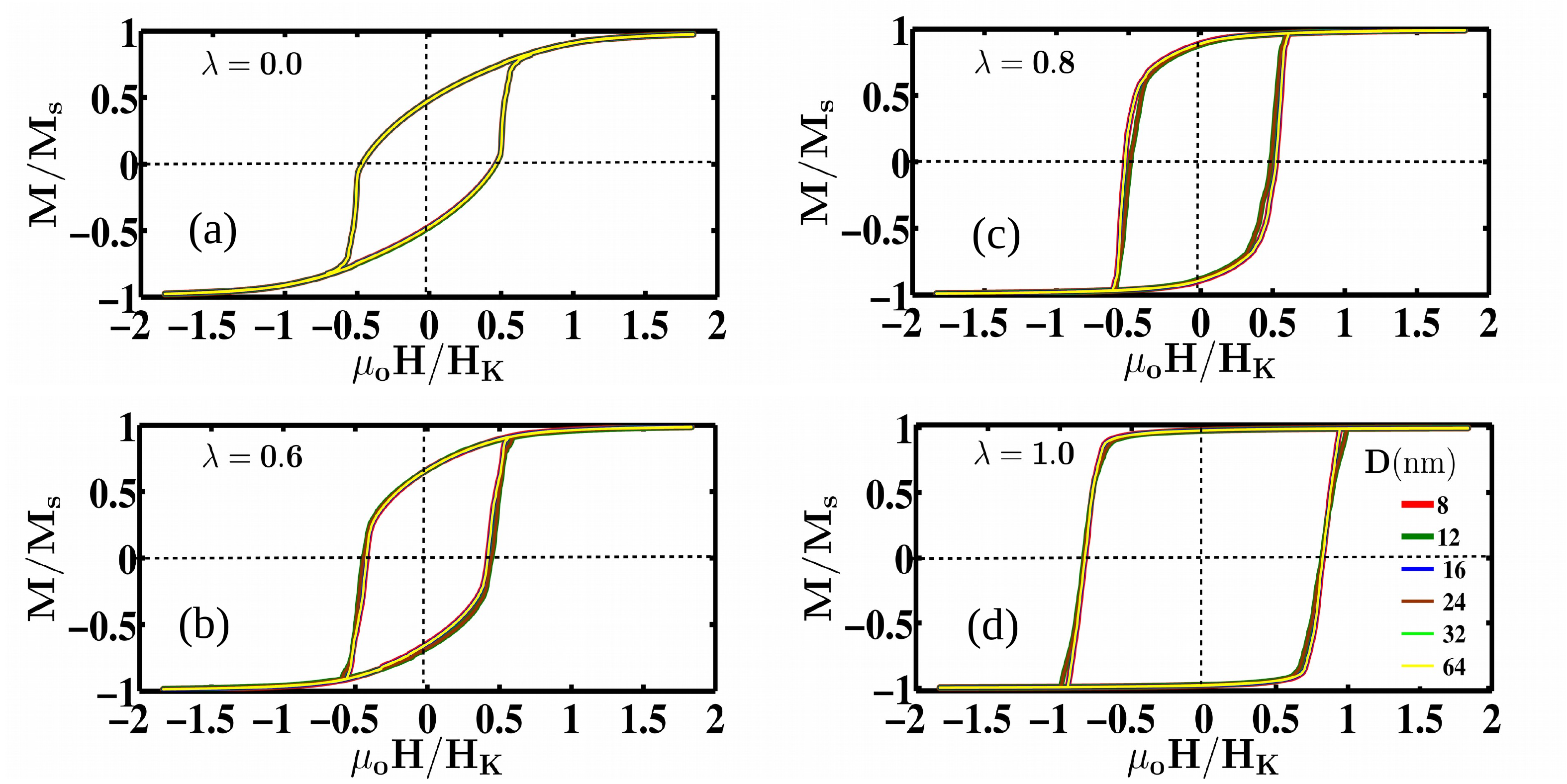}
\caption{The dependence of hysteresis curve on the particle size
$D$ for four representative values of dipolar interaction strength (a) $\lambda=0.0$, (b) $\lambda=0.6$, (c) $\lambda=0.8$ and (d) $\lambda=1.0$ at temperature $T=300$ K. The hysteresis curve follows the Stoner and Wohlfarth model for $\lambda=0.0$, irrespective of $D$. There is an increase in hysteresis loop area with dipolar interaction strength $\lambda$ due to an increase in ferromagnetic coupling. 
} 
\label{figure2}
\end{figure}

Next, we study the dependence of magnetic hysteresis properties on dipolar interaction strength for the particle sizes varied from superparamagnetic to  the ferromagnetic regime.
In Fig.~(\ref{figure2}), we analysed the hysteresis properties with particle size $D$ four typical values of $\lambda=0.0$, 0.6, 0.8 and 1.0 in the absence of temperature. We have considered six representative values of $D=8$, 12, 16, 24, 32 and 64 nm. The profile of the hysteresis curve is the same as that of Stoner and Wohlfarth particle for $T=0$ K and $\lambda=0.0$, as expected~\cite{stoner1948}. As thermal fluctuations are absent in the present case, the system behaves like a blocked state, irrespective of $D$. Therefore, the hysteresis loop area is significant even for superparamagnetic nanoparticle with negligible dipolar interaction. The hysteresis loop area also does not depend on $D$ for a fixed value of $\lambda$. The dipolar interaction of enough strength induces ferromagnetic coupling between the MNPs. Therefore, the area under the hysteresis curve increases with dipolar interaction $\lambda$.

\begin{figure}[!htb]
\centering\includegraphics[scale=0.50]{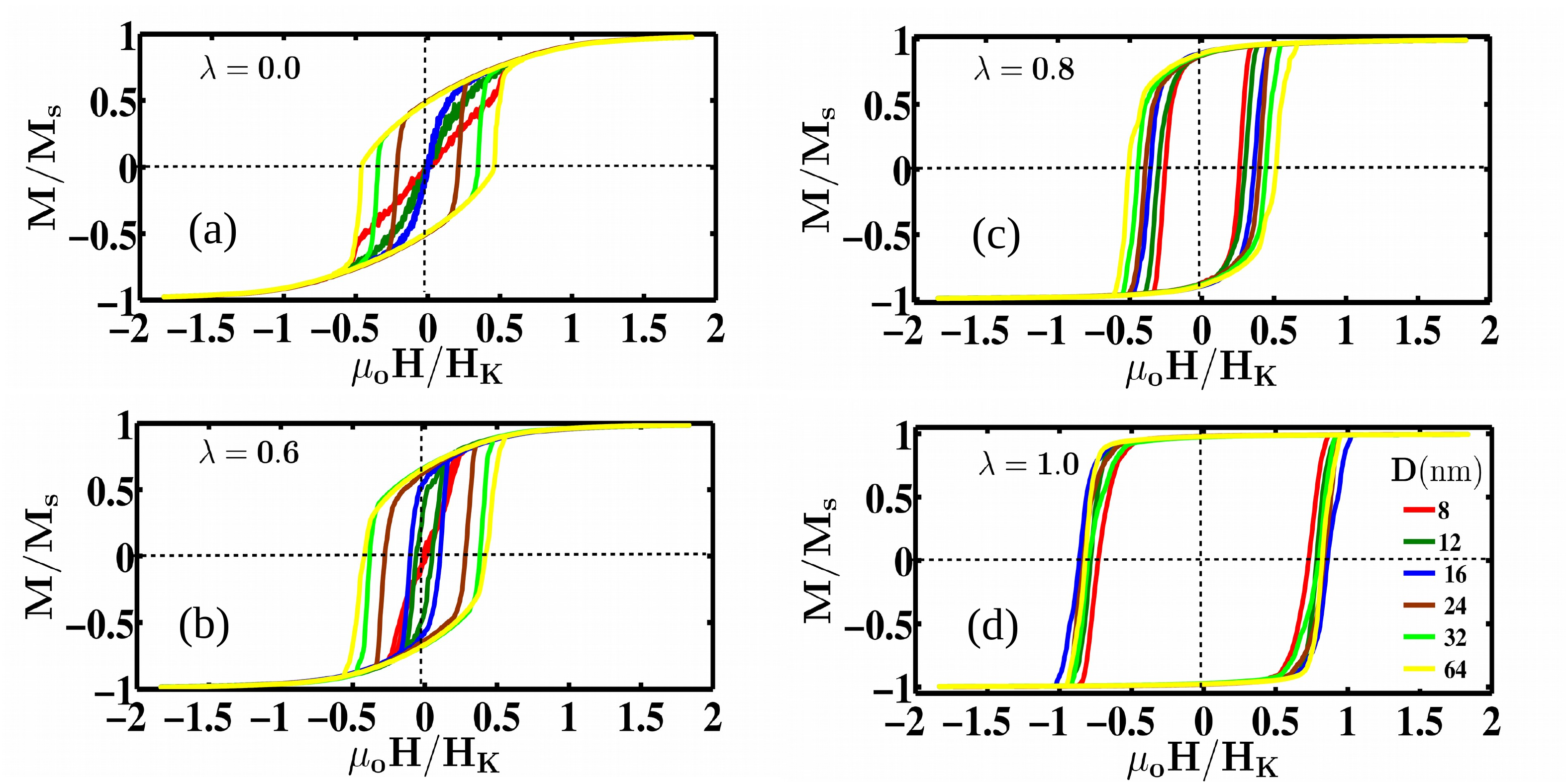}
\caption{The study of hysteresis behaviour as a function $D$ for four representative values of dipolar interaction strength (a)  $\lambda=0.0$, (b)  $\lambda=0.6$, (c)  $\lambda=0.8$ and (d)  $\lambda=1.0$ at $T=300$ K. The area under hysteresis curve is minimal for superparamagnetic nanoparticle ($D=8-16$ nm) in the non-interacting or weakly dipolar interacting case. On the other hand, for ferromagnetic nanoparticle ($D>16$ nm), the hysteresis loop area is significant and follows the Stoner and Wohlfarth model even with $T=300$ K and $\lambda=0.0$. There is an enhancement in the area under the hysteresis as particle size $D$, and dipolar coupling strength $\lambda$ is increased.}
\label{figure3}
\end{figure}

To see the thermal effects on the hysteresis for these particles, we plot the  hysteresis curves as a function of $D$ and $\lambda$ at room temperature in Fig.~(\ref{figure3}). We have used the same set of parameters as that of Fig.~(\ref{figure2}).
In the absence of dipolar interaction ($\lambda=0.0$), there is an extremely small hysteresis loop area for $D\approx8-16$ nm, indicating the dominance of superparamagnetic character. As far as the role of blocking temperature is concerned for such behaviour, $T^{}_B$ lies in the range $\approx 17-135$ K for $D=8-16$ nm, which is well below the simulation temperature $T=300$ K. As a consequence, thermal fluctuations dictate the hysteresis of superparamagnetic nanoparticles in the non-interacting case, resulting in extremely weak hysteresis. Our observation for superparamagnetic nanoparticle ($D=8$ nm) is in qualitative agreement with the numerical work of Usov and Grebenshchikov~\cite{usov2009}. We could not compare for the entire range of $D$ as they concentrated on $D=8$ nm only. So, our results are more generic in this context. The hysteresis loop area is significant for $D>16$ nm even with $\lambda=0.0$.
It means that particle with size greater than $16$ nm shows the ferromagnetic character. It is further strengthened from the fact that $T^{}_B$ is close to $300$ K for $D>16$ nm. Therefore, we observe Stoner and Wohlfarth-like behaviour even with $T=300$ K and negligible dipolar interaction ($\lambda=0.0$). Using experiments, Li {\it et al.} also obtained a large hysteresis loop area for ferromagnetic nanoparticles even in the non-interacting case~\cite{li2011}. 
We could not compare our results for the dipolar interacting nanoparticles for the whole range of particle size as they studied with a single value of particle size and $\lambda\approx0.0$ only. So, our results are more generic and can be used as a benchmark in this context. The dipolar interaction increases the hysteresis loop area as it induces ferromagnetic coupling between the MNPs. 
For the closest packing ($\lambda=1.0$), there is less dependence of hysteresis on $D$ because of huge ferromagnetic coupling as dipolar interaction is the maximum. These results suggest that ferromagnetic nanoparticles could be one of the best candidates to achieve significant heat dissipation which is essential for magnetic hyperthermia applications. 
The appreciable hysteresis loop area in the case of weak or negligible dipolar interaction with $D>16$ nm suggests that heating efficiency can be much improved with ferromagnetic nanoparticles even in the isotropic assembly of MNPs.

\begin{figure}[!htb]
\centering\includegraphics[scale=0.50]{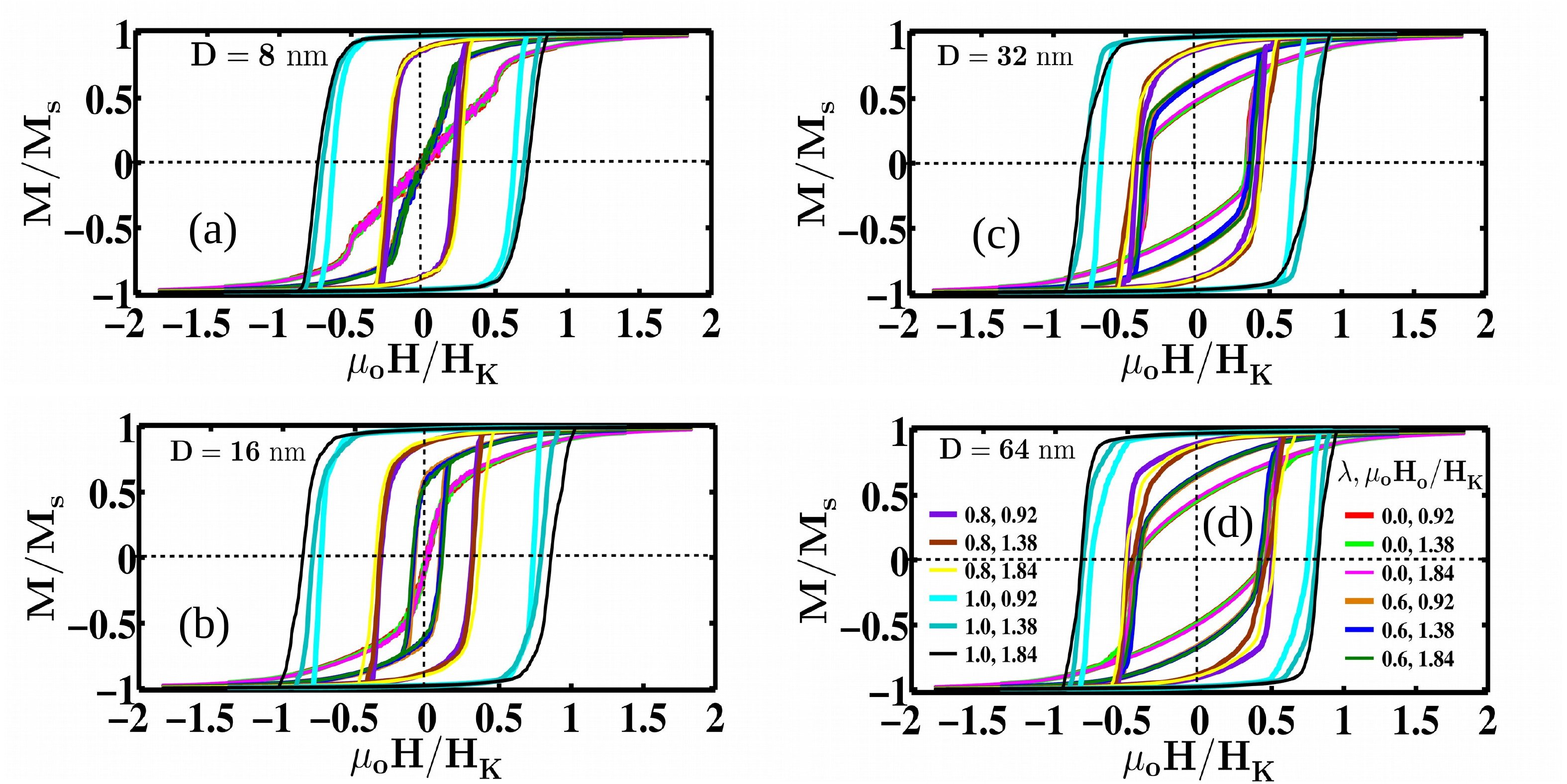}
\caption{Magnetic hysteresis curve for various values of dipolar coupling strength $\lambda$ and three values of external magnetic strength $\mu^{}_oH^{}_o$ (scaled by $H^{}_K$). We have considered four representative values of $D=$ (a) 8 nm, (b) 16 nm, (c) 32 nm and (d) 64 nm at $T=300$ K. There is an extremely small hysteresis loop area for superparamagnetic nanoparticle with negligible dipolar interaction, irrespective of magnetic field strength. When there is sufficient dipolar coupling, the hysteresis loop area increases with an increase in external field strength.}
\label{figure4}
\end{figure}

It is equally important to investigate the effect of external magnetic field strength $\mu^{}_oH^{}_o$ on the hysteresis loop area as a function of particle size. In Fig.~(\ref{figure4}), we plot the hysteresis curves for three values of $\mu^{}_oH^{}_o=0.05$, 0.075 and 0.1 T with various values of $D$, $\lambda$ at room temperature.
 In the absence of magnetic coupling ($\lambda=0.0$), there is an extremely small value of hysteresis loop area for $D\leq16$ nm irrespective of $\mu^{}_oH^{}_o$, indicating superparamagnetic behaviour. There is a significant hysteresis loop area for the ferromagnetic particle ($D>16$ nm) even for non-interacting nanoparticles, which is robust with respect to the field strength. The hysteresis loop also increases with external magnetic field strength. It is in qualitative agreement with the experimental work of Avugadda {\it et al.}~\cite{avugadda2019}. The rise in the hysteresis loop area with external magnetic field strength is also in qualitative agreement with experimental and numerical works of Kita {\it et al.}~\cite{kita2010}. There is always an increase in hysteresis loop area as dipolar interaction strength, and particle size are increased. These results also indicate that we can have considerable heat dissipation with ferromagnetic nanoparticles even in an isotropic assembly and in the absence of magnetic interaction. So, it can be used for efficient application of magnetic hyperthermia. These results should also help choose precise values of magnetic field strength, dipolar interaction, and particle size to obtain desired heat dissipation, which is essential for magnetic hyperthermia application.

\begin{figure}[!htb]
\centering\includegraphics[scale=0.50]{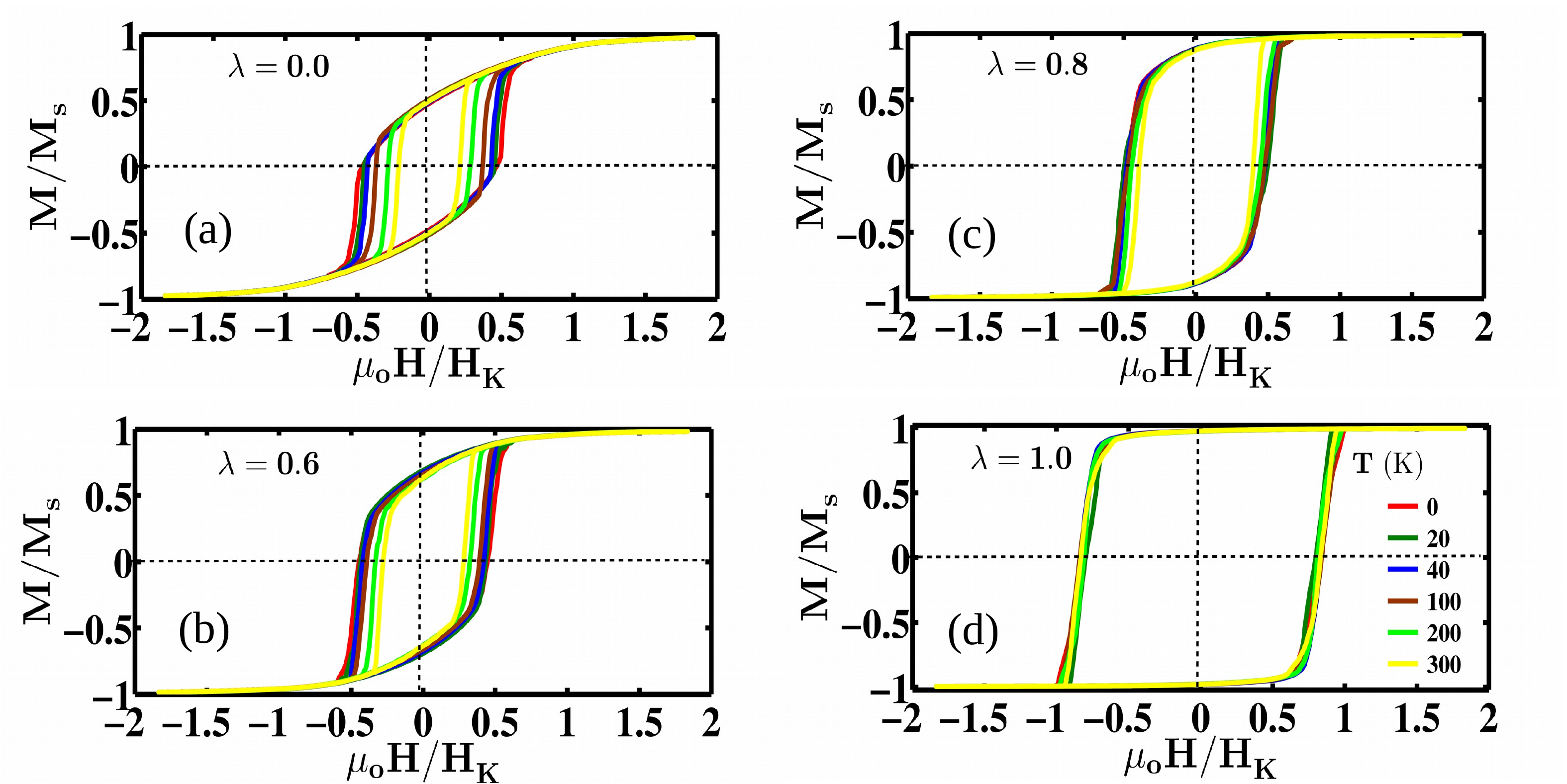}
\caption{Analysis of hysteresis loop for ferromagnetic nanoparticle ($D=24$ nm) as a function of temperature $T$ with four typical values of dipolar interaction strength $\lambda=$ (a) 0.0, (b) 0.6, (c) 0.8 and (d) 1.0. In the absence of dipolar interaction, the area under the hysteresis curve decreases with $T$. The hysteresis loop area has a negligible dependence on $T$ in the presence of considerable magnetic interaction.}
\label{figure5}
\end{figure}

To probe further, we then study the hysteresis behaviour of ferromagnetic nanoparticle as a function of temperature for various values of dipolar interaction strength. In Fig.~(\ref{figure5}), we plot hysteresis curves for $D=24$ nm with six representative values of $T= 0$, 20, 40, 100, 200 and 300 K. We have also considered four distinct values of $\lambda=0.0$, 0.6, 0.8 and 1.0. As the blocking temperature $T^{}_B\approx455$ K exceeds the maximum simulation temperature $T=300$ K considered in the present work; an appreciable hysteresis is observed for negligible dipolar interaction even at room temperature. There is a sharp decrease in the hysteresis loop area as $T$ is changed between 0 and 300 K for $\lambda\leq0.6$. The hysteresis loop area depends weakly on temperature for large $\lambda$, which can be attributed to enhanced $T^{}_B$ because of large magnetic interaction.
Due to the rise in ferromagnetic coupling with $\lambda$, there is always an increase in hysteresis loop area with dipolar interaction strength. These observations also strengthen the fact the ferromagnetic nanoparticles can have significant heating efficiency, irrespective of the shape of the assembly.

\begin{figure}[!htb]
\centering\includegraphics[scale=0.50]{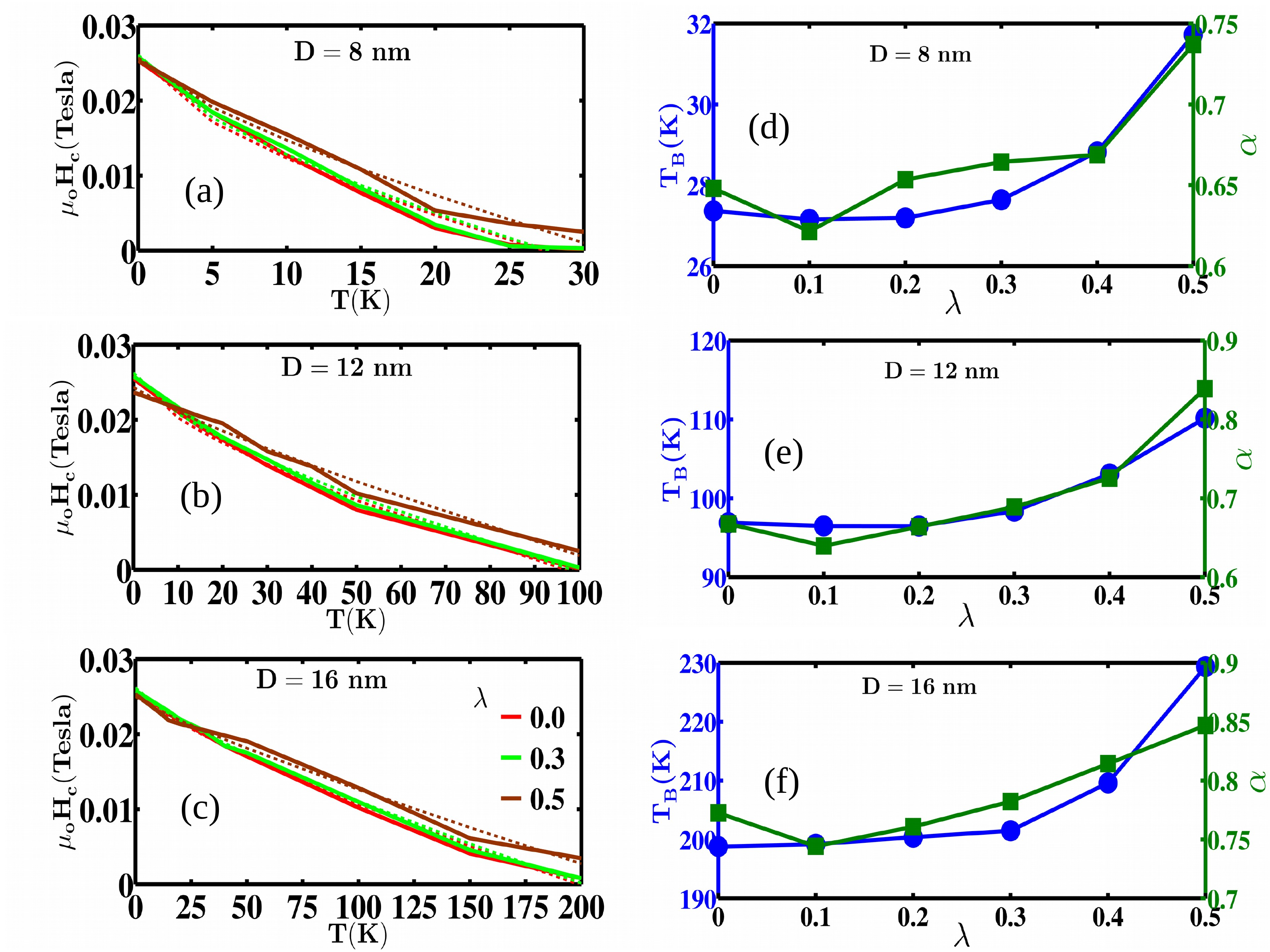}
\caption{ Coercive field $\mu^{}_oH^{}_c$ variation as a function of $T$ for three values of $\lambda=0.0$, 0.3 and 0.5 and three values of $D=$ (a) 8 nm, (b) 12 nm and (c) 16 nm. We have fitted $\mu^{}_oH^{}_c$ vs $T$ with Eq.~(\ref{coercive}) to extract $T^{}_B$ and $\alpha$. The solid line is for simulated data, and the dashed line is for the corresponding fit. In the presence of negligible dipolar interaction, $\mu^{}_oH^{}_c$ decreases rapidly with $T$. It decays very slowly with $T$ for the moderate value of $\lambda$. The extracted values of $T^{}_B$ and $\alpha$ are plotted as a function of $\lambda$ for three values of $D=$ (d) 8 nm, (e) 12 nm and (f) 16 nm. $T^{}_B$ and $\alpha$ increase with an increase in $D$ and $\lambda$.}
\label{figure6}
\end{figure}

To quantify the hysteresis mechanism, we then study the variation of coercive field $\mu_oH^{}_c$ as a function of temperature $T$, particle size $D$ and dipolar interaction strength $\lambda$. In Fig.~(\ref{figure6}), we plot simulated coercive field $\mu^{}_oH^{}_c$ as a function of $T$ for three values of $D=8, 12$ and 16 nm and three representative values of $\lambda=0.0$, 0.3 and 0.5. We have not considered larger values of $D$ ($>16$ nm), as $\mu^{}_oH^{}_c$ depends weakly on $T$ in these cases due to the dominance of ferromagnetic character for larger particle sizes. For a fixed particle size, we have also considered only those values of $T$ for which $\mu^{}_oH^{}_c$ is non-zero. It is quite evident that $\mu^{}_oH^{}_c$ decreases to zero very rapidly with $T$ for very small and negligible dipolar interaction strength. The rapid decrease of $\mu^{}_oH^{}_c$ with $T$ is in qualitative agreement with Usov and Grebenshchikov~\cite{usov2009}. The comparison is not possible for the whole range of $D$ and $\lambda$ as they have shown results only for $D=8$ nm with the dilute assembly of MNPs.
For moderate dipolar interaction strength $\lambda$, $\mu^{}_{o}H^{}_c$ decreases very slowly.

It is also quite evident that blocking temperature $T^{}_B$ is a useful quantifier in determining the dependence of hysteresis response on particle size $D$ and dipolar interaction strength $\lambda$. To extract $T^{}_B$ from the simulated hysteresis curves, we have used the model proposed by Garcia-Otero {\it et al.}~\cite{garcia1998}. In the case of non-interacting MNPs, the temperature dependence of the coercive field
$\mu^{}_oH^{}_c(T)$ can be given by the following relation~\cite{garcia1998,aquino2019} 

\begin{equation}
\mu^{}_oH^{}_c(T)=\epsilon \frac{2K^{}_{\mathrm {eff}}}{M^{}_s}\bigg[1-\bigg(\frac{T}{T^{}_B}\bigg)^{\alpha} \bigg],
\label{coercive}
\end{equation}
where $\epsilon\approx0.96$ and $\alpha=0.5$ for perfectly aligned anisotropy case while $\epsilon\approx0.479$ and $\alpha=0.75$ with randomly oriented anisotropy axes. As MNPs are interacting due to the dipolar interaction in our case, we also consider $\alpha$ as a fitting parameter. We have taken $\epsilon\approx0.479$ as anisotropy axes are assumed to have random orientations in the present work. We have fitted the simulated values of $\mu^{}_o H^{}_c$ for values as mentioned earlier of D and $\lambda$ with Eq.~(\ref{coercive}) and extracted $T^{}_B$ and $\alpha$. The obtained values of $T^{}_B$ and $\alpha$ are also plotted as a function of $\lambda$ in Fig.~(\ref{figure6}). 
The dipolar interaction and particle size $D$ enhance the blocking temperature $T^{}_B$. As a consequence, MNPs stays in the blocked state even at a substantial value of simulation temperature for superparamagnetic nanoparticle with considerable dipolar interaction strength. The same trend is observed with ferromagnetic nanoparticle and negligible dipolar interaction. 
The more significant value of $T^{}_B$ with dipolar interaction as compared to the non-interacting case is in perfect agreement with experimental works of Mehdaoui {\it et al.}~\cite{mehdaoui2010}. The increase of $T^{}_B$ with $\lambda$ is also in qualitative agreement with numerical works of Russier {\it et al.}~\cite{russier2016}. 
The remarkable thing to note is that $\alpha$ also increases with $\lambda$, and $D$. Researchers used Eq.~(\ref{coercive}) with the constant value of $\alpha$
even for strongly interacting MNPs~\cite{aquino2019,lacroix2009,enpuku2020}.
Our results suggest that $\alpha$ should not be taken as constant; instead, it should also be considered as a fitting parameter; otherwise, we may estimate incorrect values of various parameters of interest, such as $T^{}_B$ and $K^{}_{\mathrm {eff}}$.

\begin{figure}[!htb]
\centering\includegraphics[scale=0.50]{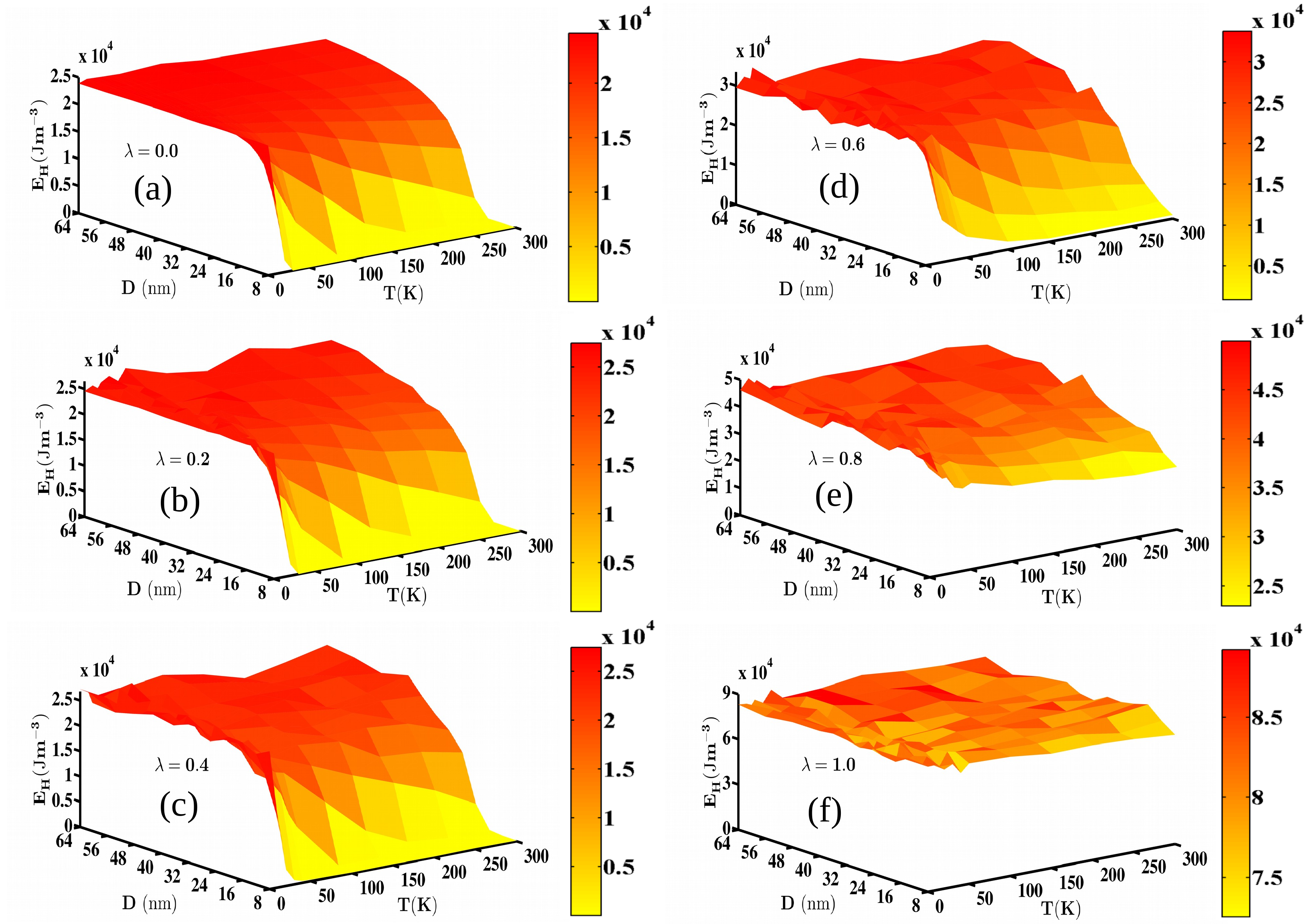}
\caption{Variation of heat dissipated due to hysteresis $E^{}_H$ as a function of particle size $D$ and temperature $T$. 
The following values of magnetic interaction strength $\lambda=$ (a) 0.0, (b) 0.2, (c) 0.4, (d) 0.6, (e) 0.8 and (f) 1.0. 
In the weak dipolar limit, $E^{}_H$ decreases rapidly with $T$ for superparamagnetic nanoparticle ($D\approx8-16$ nm). The hysteresis loop area is significant, and it also depends very weakly on $T$ for ferromagnetic nanoparticle ($D>16$ nm) even with negligible dipolar interaction. Irrespective of $D$; $E^{}_H$ depends very weakly on $T$ for large dipolar coupling.}
\label{figure7}
\end{figure}

Finally, we study the amount of heat dissipated due to the magnetic hysteresis $E^{}_H$ as a function of particle size $D$ and temperature $T$ in Fig.~(\ref{figure7}).
$D$ has been varied between 8 and 64 nm, and $T$ has been changed in the range $0-300$ K. In the case of weak and negligible dipolar interaction strength $\lambda$; $E^{}_H$ decreases very rapidly with temperature $T$ for the superparamagnetic nanoparticle ($D\approx8-16$ nm). On the other hand, in the case of ferromagnetic nanoparticle ($D>16$ nm), $E^{}_H$ is enormous, and it also depends very weakly on $T$ even in the non-interacting case. It can be explained from the fact that the blocking temperature $T^{}_B$ is comparable to or greater than the maximum simulation temperature $T=300$ K. As a consequence; the MNPs remain in the blocked state even at room temperature, which results in large hysteresis loop area and hence $E^{}_H$ is significantly high. As the dipolar interaction induces ferromagnetic coupling in an anisotropic assembly of MNPs, $T^{}_B$ increases with $\lambda$. Therefore, $E^{}_H$ increases with $\lambda$ and has a very weak dependence on $T$ as well. Zubarev {\it et al.} also found an increase in heat dissipation because of dipolar interaction with ferromagnetic nanoparticles~\cite{zubarev2018}. 
In the case of weak and negligible dipolar interaction strength, the rapid decrease of $E^{}_H$ with $T$ for superparamagnetic nanoparticle is in perfect agreement with Carrey {\it et al.}~\cite{carrey2011}. We could not compare our results in the presence of dipolar interactions as they have concentrated only on $\lambda=0.0$. Therefore, our results could be used as a benchmark in this context. The decrease of hysteresis loop area with temperature for superparamagnetic nanoparticles is also in qualitative agreement with the experimental works of Hu {\it et al.}~\cite{hu2019}. 
The coercive field $\mu^{}_oH^{}_c$ variation with $D$ and $T$ is found to be exactly same as that of $E^{}_H$ variation. So, we have not shown the curves for  $\mu^{}_oH^{}_c$ variation to avoid duplications. It is in also in perfect agreement with the work of Carrey {\it et al.} with $\lambda=0.0$~\cite{carrey2011}. These results can be used in optimizing heat dissipation by selecting precise values materials parameters such particle size, dipolar interaction strength and temperature. 


\section{Summary and Conclusion}
Now, we summarize and discuss the main results of the present work. The hysteresis curve follows the Stoner-Wohlfarth model in the absence of dipolar interaction and temperature, irrespective of particle size, as expected~\cite{stoner1948}. 
The perfect agreement between simulation and theoretical model also validates the kMC method used. In the presence of sufficient thermal energy, we observe an extremely small hysteresis loop area for the smaller particle  ($D\approx8-16$ nm) and negligible dipolar interaction, indicating the dominance of superparamagnetic character. Remarkably, the hysteresis loop area is significant with ferromagnetic particle ($D>16$ nm) and is the same as that of Stoner and Wohlfarth particle even at high value of temperature and weak or negligible dipolar interaction. Baker {\it et al.} also observed substantial hysteresis loss for ferromagnetic nanoparticle of Fe$_2$O$_3$~\cite{baker2006}. It is also in perfect agreement with the experimental works of Hergt {\it et al.}~\cite{hergt2005}. The dipolar interaction induces ferromagnetic coupling between the magnetic moments, which moves the underlying system from superparamagnetic regime to a ferromagnetic state even at sufficiently high value of temperature. Therefore, the dipolar coupling of enough strength enhances the hysteresis loop area significantly even at $T=300$ K irrespective of particle size. These observations are also in qualitative agreement with Zubarev {\it et al.}~\cite{zubarev2019}.
The hysteresis loop area is greatly hampered by the thermal fluctuations, which is quite evident with superparamagnetic particle and negligible dipolar interaction. Remarkably, the hysteresis loop area depends weakly on temperature with ferromagnetic nanoparticle even in the absence of dipolar interaction. The same trend is also observed in the case of superparamagnetic nanoparticle with sufficient dipolar interaction.

The coercive field $\mu^{}_oH^{}_c$ is found to be an essential quantifier to understand the hysteresis response of these nanosystems. In the case of weak magnetic interaction and small particle size, $\mu^{}_oH^{}_c$ decreases rapidly with temperature. It is due to fact that MNPs change their magnetic state from blocked to superparamagnetic state very rapidly with an increase in thermal fluctuations. Pauly {\it et al.} also found similar behaviour of $\mu^{}_oH^{}_c$ with $T$ using experiments~\cite{pauly2012}.
Because of significant ferromagnetic coupling due to the dipolar interaction, magnetic nanoparticles remain in the blocked state even at a tremendous 
value of temperature. Therefore, $\mu_oH^{}_c$ decreases very weakly as a function of temperature in the presence of considerable dipolar interaction strength. The same trend is observed with ferromagnetic particles, even in an assembly of non-interacting MNPs. In addition to $\mu^{}_oH^{}_c$, blocking temperature $T^{}_B$ is also a useful measure to probe the hysteresis mechanism as a function of particle size and dipolar interaction. So, we extracted $T^{}_B$ from the simulated hysteresis curves using  Garcia-Otero {\it et al.} model for the temperature dependence of the coercive field~\cite{garcia1998}. The so-obtained $T^{}_B$ increases with particle size and dipolar interaction strength. Pauly {\it et al.} also observed an enhancement in $T^{}_B$ with particle size~\cite{pauly2012}. Remarkably, $T^{}_B$ exceeds the highest simulation temperature considered in the present work ($T=300$ K) for ferromagnetic nanoparticles. Therefore, we observe the Stoner and Wohlfarth-like behaviour even at $T=300$ K and negligible dipolar interaction for such nanoparticles. In general, two distinct regimes are depending on the relative strength of $T^{}_B$ and simulation temperature $T$. For $T<T^{}_B$, the energy barriers trap the magnetic moments associated with particles in metastable orientation, resulting in a blocked state. On the contrary, the thermal energy $k^{}_BT$ overcomes the energy barrier for $T>T^{}_B$ resulting in the well-known superparamagnetic regime. The fitting of simulated values of coercive field $\mu^{}_oH^{}_c$ with Garcia-Otero {\it et al.} model also suggests that $T^\alpha$ dependence of coercive field with $\alpha=0.75$ deviates significantly with particle size and dipolar interaction strength. The exponent $\alpha$ is found to increase with an increase in $\lambda$ and $D$. So, $\alpha$ should not be considered constant; otherwise, it may estimate incorrect values of $T^{}_B$ and $K^{}_{\mathrm {eff}}$.

To conclude, we have systematically analysed the hysteresis mechanism in ordered arrays of magnetic nanoparticles as a function of particle size extensively.
We have also investigated the dependence of hysteresis response on dipolar interaction and temperature. Our results suggest that hysteresis response of the underlying system depends critically on these parameters. Our results strengthen the fact that ferromagnetic nanoparticles can be one of the best candidates for magnetic hyperthermia irrespective of the shape of the assembly. 
In these contexts; we have also explained a variety of experimental studies.
So, the observations made in the present work should also help experimentalist in choosing the suitable value of various parameters such as particle size, dipolar interaction strength, the magnetic field to obtain desired heating for different applications such as magnetic hyperthermia, drug delivery, etc. It should also provide a better understanding of magnetic hyperthermia to researchers working on this subject. Therefore, we believe that the present work should stimulate joint efforts in computational, experimental and analytical research for these precious nano heaters.


\bibliographystyle{h-physrev}
\bibliography{ref}

\begin{thebibliography}{10}

\bibitem{bader2006}
S.~D. Bader,
\newblock Reviews of Modern Physics {\bf 78}, 1 (2006).

\bibitem{gloag2019}
L.~Gloag, M.~Mehdipour, D.~Chen, R.~D. Tilley, and J.~J. Gooding,
\newblock Advanced Materials {\bf 31}, 1904385 (2019).

\bibitem{koplovitz2019}
G.~Koplovitz, G.~Leitus, S.~Ghosh, B.~P. Bloom, S.~Yochelis, D.~Rotem,
  F.~Vischio, M.~Striccoli, E.~Fanizza, R.~Naaman, {\em et~al.},
\newblock Small {\bf 15}, 1804557 (2019).

\bibitem{akbarzadeh2012}
A.~Akbarzadeh, M.~Samiei, and S.~Davaran,
\newblock Nanoscale Research Letters {\bf 7}, 144 (2012).

\bibitem{guimaraes2009}

\newblock A.~P. Guimar{\~a}es and A.~P. Guimaraes{\em Principles of
  Nanomagnetism} Vol.~7 (Springer, 2009).

\bibitem{gu2016}
M.~Gu, Q.~Zhang, and S.~Lamon,
\newblock Nature Reviews Materials {\bf 1}, 1 (2016).

\bibitem{patra2018}
J.~K. Patra, G.~Das, L.~F. Fraceto, E.~V.~R. Campos, M.~del Pilar
  Rodriguez-Torres, L.~S. Acosta-Torres, L.~A. Diaz-Torres, R.~Grillo, M.~K.
  Swamy, S.~Sharma, {\em et~al.},
\newblock Journal of Nanobiotechnology {\bf 16}, 71 (2018).

\bibitem{anand2018}
M.~Anand, J.~Carrey, and V.~Banerjee,
\newblock Journal of Magnetism and Magnetic Materials {\bf 454}, 23 (2018).

\bibitem{Aldewachi2018}
H.~Aldewachi, T.~Chalati, M.~N. Woodroofe, N.~Bricklebank, B.~Sharrack, and
  P.~Gardiner,
\newblock Nanoscale {\bf 10}, 18 (2018).

\bibitem{anand12021}
M.~Anand,
\newblock arXiv preprint arXiv:2102.13440  (2021).

\bibitem{berry2019}
S.~L. Berry, K.~Walker, C.~Hoskins, N.~D. Telling, and H.~P. Price,
\newblock Scientific Reports {\bf 9}, 1 (2019).

\bibitem{moise2018}
S.~Moise, J.~M. Byrne, A.~J. El~Haj, and N.~D. Telling,
\newblock Nanoscale {\bf 10}, 20519 (2018).

\bibitem{martinez2013}
C.~Martinez-Boubeta, K.~Simeonidis, A.~Makridis, M.~Angelakeris, O.~Iglesias,
  P.~Guardia, A.~Cabot, L.~Yedra, S.~Estrad{\'e}, F.~Peir{\'o}, {\em et~al.},
\newblock Scientific Reports {\bf 3}, 1652 (2013).

\bibitem{perigo2015}
E.~A. Perigo, G.~Hemery, O.~Sandre, D.~Ortega, E.~Garaio, F.~Plazaola, and
  F.~J. Teran,
\newblock Applied Physics Reviews {\bf 2}, 041302 (2015).

\bibitem{abu2020}
A.~F. Abu-Bakr and A.~Y. Zubarev,
\newblock Philosophical Transactions of the Royal Society A {\bf 378}, 20190251
  (2020).

\bibitem{rosensweig2002}
R.~E. Rosensweig,
\newblock Journal of Magnetism and Magnetic Materials {\bf 252}, 370 (2002).

\bibitem{raikher2014}
Y.~L. Raikher and V.~Stepanov,
\newblock Journal of Magnetism and Magnetic Materials {\bf 368}, 421 (2014).

\bibitem{ruta2015}
S.~Ruta, R.~Chantrell, and O.~Hovorka,
\newblock Scientific Reports {\bf 5}, 9090 (2015).

\bibitem{de1997}
K.~De'Bell, A.~B. MacIsaac, I.~N. Booth, and J.~P. Whitehead,
\newblock Phys. Rev. B {\bf 55}, 15108 (1997).

\bibitem{iglesias2004}
{\`O}.~Iglesias and A.~Labarta,
\newblock Physical Review B {\bf 70}, 144401 (2004).

\bibitem{plumer2010}
M.~Plumer, J.~van Lierop, B.~Southern, and J.~Whitehead,
\newblock Journal of Physics: Condensed Matter {\bf 22}, 296007 (2010).

\bibitem{anand2016}
M.~Anand, J.~Carrey, and V.~Banerjee,
\newblock Physical Review B {\bf 94}, 094425 (2016).

\bibitem{usov2018}
N.~Usov, M.~Nesmeyanov, and V.~Tarasov,
\newblock Scientific Reports {\bf 8}, 1 (2018).

\bibitem{anand2021}
M.~Anand,
\newblock arXiv preprint arXiv:2101.03356  (2021).

\bibitem{djurberg1997}
C.~Djurberg, P.~Svedlindh, P.~Nordblad, M.~F. Hansen, F.~B{\o}dker, and
  S.~M{\o}rup,
\newblock Physical Review Letters {\bf 79}, 5154 (1997).

\bibitem{barrett2011}
M.~Barrett, A.~Deschner, J.~P. Embs, and M.~C. Rheinst{\"a}dter,
\newblock Soft Matter {\bf 7}, 6678 (2011).

\bibitem{vargas2005}
J.~M. Vargas, W.~C. Nunes, L.~M. Socolovsky, M.~Knobel, and D.~Zanchet,
\newblock Phys. Rev. B {\bf 72}, 184428 (2005).

\bibitem{souza2019}
C.~M. Souza, S.~S. Pedrosa, A.~S. Carri\ifmmode~\mbox{\c{c}}\else \c{c}\fi{}o,
  G.~O.~G. Rebou\ifmmode~\mbox{\c{c}}\else \c{c}\fi{}as, and A.~L. Dantas,
\newblock Phys. Rev. B {\bf 99}, 174441 (2019).

\bibitem{allia1999}
P.~Allia, M.~Coisson, M.~Knobel, P.~Tiberto, and F.~Vinai,
\newblock Physical Review B {\bf 60}, 12207 (1999).

\bibitem{tan2014}
R.~P. Tan, J.~Carrey, and M.~Respaud,
\newblock Phys. Rev. B {\bf 90}, 214421 (2014).

\bibitem{fu2018}
R.~Fu, Y.~Yan, C.~Roberts, Z.~Liu, and Y.~Chen,
\newblock Scientific Reports {\bf 8}, 1 (2018).

\bibitem{ha2018}
Y.~Ha, S.~Ko, I.~Kim, Y.~Huang, K.~Mohanty, C.~Huh, and J.~A. Maynard,
\newblock ACS Applied Nano Materials {\bf 1}, 512 (2018).

\bibitem{bao2011}
N.~Bao and A.~Gupta,
\newblock Journal of Materials Research {\bf 26}, 111 (2011).

\bibitem{chantrell1982}
R.~Chantrell, A.~Bradbury, J.~Popplewell, and S.~Charles,
\newblock Journal of Applied Physics {\bf 53}, 2742 (1982).

\bibitem{lima2012}
E.~Lima~Jr, E.~De~Biasi, M.~V. Mansilla, M.~E. Saleta, M.~Granada, H.~E.
  Troiani, F.~Effenberger, L.~Rossi, H.~Rechenberg, and R.~D. Zysler,
\newblock Journal of Physics D: Applied Physics {\bf 46}, 045002 (2012).

\bibitem{branquinho2013}
L.~C. Branquinho, M.~S. Carri{\~a}o, A.~S. Costa, N.~Zufelato, M.~H. Sousa,
  R.~Miotto, R.~Ivkov, and A.~F. Bakuzis,
\newblock Scientific Reports {\bf 3}, 2887 (2013cs).

\bibitem{usov2017}
N.~Usov, O.~Serebryakova, and V.~Tarasov,
\newblock Nanoscale Research Letters {\bf 12}, 1 (2017).

\bibitem{guibert2015}
C.~Guibert, V.~Dupuis, V.~Peyre, and J.~Fresnais,
\newblock The Journal of Physical Chemistry C {\bf 119}, 28148 (2015).

\bibitem{serantes2014}
D.~Serantes, K.~Simeonidis, M.~Angelakeris, O.~Chubykalo-Fesenko, M.~Marciello,
  M.~D.~P. Morales, D.~Baldomir, and C.~Martinez-Boubeta,
\newblock The Journal of Physical Chemistry C {\bf 118}, 5927 (2014).

\bibitem{valdes2020}
D.~P. Vald{\'e}s, E.~Lima~Jr, R.~D. Zysler, and E.~De~Biasi,
\newblock Physical Review Applied {\bf 14}, 014023 (2020).

\bibitem{mehdaoui2013}
B.~Mehdaoui, R.~P. Tan, A.~Meffre, J.~Carrey, S.~Lachaize, B.~Chaudret, and
  M.~Respaud,
\newblock Phys. Rev. B {\bf 87}, 174419 (2013).

\bibitem{anand2020}
M.~Anand,
\newblock Journal of Applied Physics {\bf 128}, 023903 (2020).

\bibitem{anderson2016}
N.~Anderson, D.~Louie, D.~Serantes, and K.~Livesey,
\newblock arXiv preprint arXiv:1610.07678  (2016).

\bibitem{chuan2012}
E.~W. Chuan~Lim and R.~Feng,
\newblock The Journal of Chemical Physics {\bf 136}, 124109 (2012).

\bibitem{kita2008}
E.~Kita, H.~Yanagihara, S.~Hashimoto, K.~Yamada, T.~Oda, M.~Kishimoto, and
  A.~Tasaki,
\newblock IEEE Transactions on Magnetics {\bf 44}, 4452 (2008).

\bibitem{abu2019}
A.~F. Abu-Bakr and A.~Zubarev,
\newblock Philosophical Transactions of the Royal Society A {\bf 377}, 20180216
  (2019).

\bibitem{mehdaoui2011}
B.~Mehdaoui, A.~Meffre, J.~Carrey, S.~Lachaize, L.-M. Lacroix, M.~Gougeon,
  B.~Chaudret, and M.~Respaud,
\newblock Advanced Functional Materials {\bf 21}, 4573 (2011).

\bibitem{carrey2011}
J.~Carrey, B.~Mehdaoui, and M.~Respaud,
\newblock Journal of Applied Physics {\bf 109}, 083921 (2011).

\bibitem{le2016}
X.~L.~L. Le~Chang, Y.~Q.~M. Dai Di~Fan, H.~P.~M. Huan~Zhang, P.~M. Qiu
  Ying~Liu, Y.~E.~L. Wei Ming~Xue, {\em et~al.},
\newblock International Journal of Nanomedicine {\bf 11}, 1175 (2016).

\bibitem{anand2019}
M.~Anand, V.~Banerjee, and J.~Carrey,
\newblock Physical Review B {\bf 99}, 024402 (2019).

\bibitem{manish2020}
M.~Anand,
\newblock Journal of Magnetism and Magnetic Materials {\bf 522}, 167538 (2021).

\bibitem{bedanta2008}
S.~Bedanta and W.~Kleemann,
\newblock Journal of Physics D: Applied Physics {\bf 42}, 013001 (2008).

\bibitem{he2018}
S.~He, H.~Zhang, Y.~Liu, F.~Sun, X.~Yu, X.~Li, L.~Zhang, L.~Wang, K.~Mao,
  G.~Wang, {\em et~al.},
\newblock Small {\bf 14}, 1800135 (2018).

\bibitem{hachani2017}
R.~Hachani, M.~A. Birchall, M.~W. Lowdell, G.~Kasparis, L.~D. Tung, B.~B.
  Manshian, S.~J. Soenen, W.~Gsell, U.~Himmelreich, C.~A. Gharagouzloo, {\em
  et~al.},
\newblock Scientific Reports {\bf 7}, 1 (2017).

\bibitem{stoner1948}
E.~C. Stoner and E.~Wohlfarth,
\newblock Philosophical Transactions of the Royal Society of London. Series A,
  Mathematical and Physical Sciences {\bf 240}, 599 (1948).

\bibitem{usov2009}
N.~Usov and Y.~B. Grebenshchikov,
\newblock Journal of Applied Physics {\bf 106}, 023917 (2009).

\bibitem{mohapatra2020}
J.~Mohapatra, M.~Xing, J.~Beatty, J.~Elkins, T.~Seda, S.~R. Mishra, and J.~P.
  Liu,
\newblock Nanotechnology {\bf 31}, 275706 (2020).

\bibitem{li2011}
C.~H. Li, P.~Hodgins, and G.~Peterson,
\newblock Journal of Applied Physics {\bf 110}, 054303 (2011).

\bibitem{avugadda2019}
S.~K. Avugadda, M.~E. Materia, R.~Nigmatullin, D.~Cabrera, R.~Marotta, T.~F.
  Cabada, E.~Marcello, S.~Nitti, E.~J. Art{\'e}s-Iba{\~n}ez, P.~Basnett, {\em
  et~al.},
\newblock Chemistry of Materials {\bf 31}, 5450 (2019).

\bibitem{kita2010}
E.~Kita, T.~Oda, T.~Kayano, S.~Sato, M.~Minagawa, H.~Yanagihara, M.~Kishimoto,
  C.~Mitsumata, S.~Hashimoto, K.~Yamada, {\em et~al.},
\newblock Journal of Physics D: Applied Physics {\bf 43}, 474011 (2010).

\bibitem{garcia1998}
J.~Garc{\i}a-Otero, A.~Garc{\i}a-Bastida, and J.~Rivas,
\newblock Journal of Magnetism and Magnetic Materials {\bf 189}, 377 (1998).

\bibitem{aquino2019}
V.~R. Aquino, M.~Vin\'{i}cius-Ara\'{u}jo, N.~Shrivastava, M.~H. Sousa, J.~A.~H.
  Coaquira, and A.~F. Bakuzis,
\newblock The Journal of Physical Chemistry C {\bf 123}, 27725 (2019).

\bibitem{mehdaoui2010}
B.~Mehdaoui, A.~Meffre, L.-M. Lacroix, J.~Carrey, S.~Lachaize, M.~Respaud,
  M.~Gougeon, and B.~Chaudret,
\newblock Journal of Applied Physics {\bf 107}, 09A324 (2010).

\bibitem{russier2016}
V.~Russier,
\newblock Journal of Magnetism and Magnetic Materials {\bf 409}, 50 (2016).

\bibitem{lacroix2009}
L.-M. Lacroix, R.~B. Malaki, J.~Carrey, S.~Lachaize, M.~Respaud, G.~Goya, and
  B.~Chaudret,
\newblock Journal of Applied Physics {\bf 105}, 023911 (2009).

\bibitem{enpuku2020}
K.~Enpuku, A.~L. Elrefai, T.~Yoshida, T.~Kahmann, J.~Zhong, T.~Viereck, and
  F.~Ludwig,
\newblock Journal of Applied Physics {\bf 127}, 133903 (2020).

\bibitem{zubarev2018}
A.~Y. Zubarev,
\newblock Physical Review E {\bf 98}, 032610 (2018).

\bibitem{hu2019}
P.~Hu, T.~Chang, W.-J. Chen, J.~Deng, S.-L. Li, Y.-G. Zuo, L.~Kang, F.~Yang,
  M.~Hostetter, and A.~A. Volinsky,
\newblock Journal of Alloys and Compounds {\bf 773}, 605 (2019).

\bibitem{baker2006}
I.~Baker, Q.~Zeng, W.~Li, and C.~R. Sullivan,
\newblock Journal of Applied Physics {\bf 99}, 08H106 (2006).

\bibitem{hergt2005}
R.~Hergt, R.~Hiergeist, M.~Zeisberger, D.~Sch{\"u}ler, U.~Heyen, I.~Hilger, and
  W.~A. Kaiser,
\newblock Journal of Magnetism and Magnetic Materials {\bf 293}, 80 (2005).

\bibitem{zubarev2019}
A.~Y. Zubarev,
\newblock Physical Review E {\bf 99}, 062609 (2019).

\bibitem{pauly2012}
M.~Pauly, B.~P. Pichon, P.~Panissod, S.~Fleutot, P.~Rodriguez, M.~Drillon, and
  S.~Begin-Colin,
\newblock Journal of Materials Chemistry {\bf 22}, 6343 (2012).

\end{thebibliography}


\newpage

\newpage

\newpage

\newpage
\newpage

\newpage

\newpage

\end{document}